\def\rfr#1{eq. (\ref{#1})}

\def\dert#1#2{\frac{{{d}}{#1}}{{{d}}{#2}}}              

\def\bar{\begin{eqnarray}}
\def\ear{\end{eqnarray}}
\def\bb{\bibitem}
\def\eqi{\begin{equation}}
\def\eqf{\end{equation}}
\def\eqia{\begin{eqnarray}}
\def\eqfa{\end{eqnarray}}
\def\rp#1#2{{#1\over#2}}

\def\lb#1{\label{#1}}

\def\oc2{$\mathcal{O}(c^{-2})$}

\def\bds#1{\boldsymbol{#1}}

\documentclass{aastex}
\usepackage{url}
\usepackage{rotating,amsmath,amsthm,amscd,amssymb}
\usepackage{latexsym}
\usepackage{graphicx,epsfig}

\RequirePackage{color}

\begin{document}

\title{Juno, the angular momentum of Jupiter and the Lense-Thirring effect}

\shorttitle{Juno, the Lense-Thirring effect and the angular momentum of Jupiter}
\shortauthors{L. Iorio }

\author{Lorenzo Iorio }
\affil{INFN-Sezione di Pisa. Permanent address for correspondence: Viale Unit\`{a} di Italia 68, 70125, Bari (BA), Italy. E-mail: lorenzo.iorio@libero.it}

\begin{abstract}
The recently approved Juno mission will orbit Jupiter for one year in a highly eccentric ($r_{\rm min}=1.06R_{\rm Jup}$, $r_{\rm max}=39R_{\rm Jup}$) polar orbit ($i=90$ deg) to accurately map, among other things, the jovian magnetic and  gravitational fields. Such an orbital configuration yields an ideal  situation, in principle, to attempt a measurement of the general relativistic Lense-Thirring effect through the Juno's node $\Omega$ which would be displaced by about 570 m over the mission's duration. Conversely, by assuming the validity of general relativity, the proposed test can be viewed as a direct, dynamical measurement of the Jupiter's angular momentum $S$ which would give important information concerning the internal structure and formation of the giant planet. The long-period orbital perturbations due to the zonal harmonic coefficients $J_{\ell}, \ell=2,3,4,6$ of the multipolar expansion of the jovian gravitational potential accounting for its departures from spherical symmetry are, in principle, a major source of systematic bias. While the Lense-Thirring node rate is independent of the inclination $i$, the node zonal perturbations vanish for $i=90$. In reality, the orbit injection errors will induce departures $\delta i$ from the ideal
polar geometry, so that, according to a conservative analytical analysis, the zonal perturbations may come into play at an unacceptably high level, in spite of the expected improvements in the low-degree zonals by Juno. A linear combination of $\Omega$, the periJove $\omega$ and the mean anomaly $\mathcal{M}$ cancels out the impact of $J_2$ and $J_6$. A two orders of magnitude improvement in the uncanceled $J_3$ and $J_4$ would be needed to reduce their bias on the relativistic signal to the percent level; it does not seem unrealistic  because the expected level of improvement in such zonals is three orders of magnitude. More favorable conclusions are obtained by looking at single Doppler range-rate measurements taken around the closest approaches to Jupiter;  numerical simulations of the classical and gravito-magnetic signals for this kind of observable show that a $0.2-5\%$ accuracy would be a realistic goal.
\end{abstract}


\section{Introduction}

Recently, NASA approved the Juno\footnote{See on the WEB http://juno.wisc.edu/index.html} mission \citep{Mat05} to Jupiter. Juno is a spinning, solar powered spacecraft to be placed in a highly eccentric polar orbit around Jupiter (see Table \ref{tavola} for its relevant orbital parameters)
specifically designed to avoid its highest radiation regions. Understanding the formation, evolution and
structure of Jupiter is the primary science goal of
Juno.
\begin{table}[!h]
\caption{\footnotesize{Planetocentric nominal orbital parameters of Juno. $a,e,i$ are the semi-major axis (in jovian radii $R=71492$ km), the eccentricity and the inclination (in deg) to the Jupiter's equator, respectively. $P$ is the orbital period (in days). $T$ is the mission duration (in years).} }\label{tavola}
\centering
\bigskip
\begin{tabular}{ccccc}
\hline\noalign{\smallskip}
$a$ ($R$) & $e$ & $i$ (deg) & $P$ (d) & $T$ (yr)\\
\noalign{\smallskip}\hline\noalign{\smallskip}
20.03 & 0.947 & 90  & 11 & 1\\
\noalign{\smallskip}\hline
\end{tabular}
\end{table}
It will carry onboard a dual frequency gravity/radio
science system, a six wavelength microwave radiometer for atmospheric sounding and composition, a
dual-technique magnetometer, plasma detectors, energetic particle detectors, a radio/plasma wave
experiment, and an ultraviolet imager/spectrometer. The nominal mission's lifetime is 1 year. Juno is aimed, among other things, at accurately mapping the  gravitational field of Jupiter \citep{And76} with unprecedented accuracy \citep{And04} by exploiting the slow apsidal precession of its 11-day orbit.

In this paper we wish to explore the possibility offered by Juno to perform a test of general relativity by directly measuring the gravito-magnetic Lense-Thirring effect; its basics are reviewed below. Even putting aside the more or less successful attempts so far performed with other natural and artificial test particles orbiting different central bodies, it must be recalled that a satisfactorily empirical corroboration of a fundamental theory like general relativity requires that as
many independent experiments as possible are conducted by different scientists in
different laboratories; thus, it is worthwhile to try to use different gravitational
fields to perform such a test of intrinsic gravito-magnetism. Conversely, since, as we will see, the Lense-Thirring precessions are due to the proper angular momentum $\bds S$ of the orbited central body, one may also assume the existence of the general relativistic gravito-magnetism and consider such a test as a direct, dynamical measurement of the Jupiter's angular momentum through the Lense-Thirring effect; this would yield further, important information concerning the interior of Jupiter.  Indeed, the moment of inertia ratio $C/MR^2$ entering $S$ is a measure of the concentration of mass towards the center of the planet \citep{Irw03}. Such a figure, together with the measured values of the zonal\footnote{They preserve the axial symmetry.} coefficients of the gravity field accounting for its deviations from spherical symmetry may be fitted with internal models that model how the density, pressure, temperature and composition vary with depth  \citep{Irw03,Gui05,Hor08}.
Moreover, a dynamical, model-independent determination of $S$ would be important also for a better knowledge of the  history and  formation of Jupiter \citep{Mac08}.

Let us, now, briefly review the basics of the Lense-Thirring effect \citep{LT,Zel,Sof89}. In its weak-field and slow-motion approximation, the field equations of general relativity get linearized looking like those of the Maxwellian electromagnetism. Analogously with the magnetic field generated by moving electric charges, mass-energy currents give rise to a gravito-magnetic field $\bds B_{\rm g}$  \citep{MashNOVA}; far from an isolated spinning body of mass $M$  and proper angular momentum $\bds S$ it is  \citep{Lic}
\eqi \bds B_{\rm g} = -\rp{G}{c r^3}\left[\bds S -3\left(\bds S\bds\cdot\bds{\hat{r}}\right)\bds{\hat{r}}\right],\lb{BGM}\eqf   where $G$ is the Newtonian gravitational constant and $c$ is the speed of light in vacuum.
$\bds B_{\rm g}$ exerts  the non-central Lorentz-like acceleration  \citep{Sof89,MashNOVA}
 \eqi \bds A^{\rm GM} = -\rp{2}{c}{\bds v}\bds\times\bds B_{\rm g}\lb{acc_gm}\eqf
 upon a test particle moving with velocity $\bds v$.
 For ordinary astronomical bodies like, e.g., a planet, $A^{\rm GM}$  is many orders of magnitude smaller than the Newtonian monopole $A^{\rm N} = {GM}/{r^2}$, so that it can be considered as a small perturbation.
As a result, the longitude of the ascending node $\Omega$ and the argument of pericentre $\omega$ of a test particle undergo small secular precessions  \citep{LT}
\begin{eqnarray}
  \dot\Omega_{\rm LT} &=& \rp{2GS}{c^2 a^3(1-e^2)^{3/2}}, \\
  \nonumber\\
  \dot\omega_{\rm LT} &=& -\rp{6GS\cos i}{c^2 a^3(1-e^2)^{3/2}}.
\end{eqnarray}
Concerning a direct  measurement of the Lense-Thirring effect, attempts have been recently performed with the LAGEOS satellites in the gravitational field of the Earth \citep{Cug,Ciu04,Ries08}, the Mars Global Surveyor probe orbiting Mars \citep{MGS,IorCEJP2} and some of the inner planets of the Solar System \citep{VENLT}. The evaluation of the total accuracy  of the LAGEOS \citep{Ciumerda,Iorioreply} and MGS \citep{LTkrogh,IorCEJP2} tests  has raised a debate in the recent past because of the difficulty of realistically assessing the impact of certain competing dynamical effects acting as sources of systematic errors; for example, the total accuracy of the LAGEOS test may be as large as some tens percent \citep{IorCEJP1}; similar shortcomings may affect also the approved LARES mission \citep{ISSI08,ASR}. For an overview of several theoretical and experimental features of gravito-magnetism see, e.g., \citep{IorNOVA}.
Concerning the jovian scenario, \citet{LT} originally proposed to use the orbital precessions of the Galilean satellites; such a possibility has been recently investigated by \citet{IorJup}, but it seems to be still premature. \citet{Haas} proposed a spacecraft-based experiment in the gravitational field of Jupiter to measure another gravito-magnetic effect, i.e. the precession of a gyroscope \citep{Pugh,Schi}. It is also one of the goals of the GP-B mission \citep{Eve} whose target was a $\approx 1\%$ measurement of such an effect with four superconducting gyroscopes carried onboard by a low-altitude polar spacecraft in the gravitational field of the Earth, but it is still unclear if it will be finally possible to meet the original accuracy because of some unexpected systematic aliasing effects occurred during the mission\footnote{See on the WEB http://einstein.stanford.edu/} \citep{Conk,Eve09}.
A test of gravito-magnetism\footnote{In this case, the mass currents inducing a gravito-magnetic action are not those related to the Jupiter's proper rotation (intrinsic gravito-magnetism), but are due to its translational orbital motion (extrinsic gravito-magnetism).} concerning the deflection of electromagnetic waves by Jupiter in its orbital motion has been performed in a dedicated radio-interferometric experiment \citep{Fom08}.
  With regard to other suggested non-gravito-magnetic  tests of general relativity in the jovian gravitational field, \citet{His} proposed to measure the much larger gravito-electric Einstein pericenter precessions \citep{Ein} of the natural satellites of Jupiter and Saturn.  There exist also plans for performing a test of the light bending due to the Jupiter's monopole and quadrupole mass moments with the forthcoming astrometric mission GAIA \citep{Cro}.

The Jupiter's proper angular momentum amounts to  \citep{Sof03}
\eqi S \approx 6.9\times 10^{38}\ {\rm kg\ m^2\ s^{-1}}.\lb{LT}\eqf
Table \ref{tavola} and \rfr{LT}
yield for Juno
\begin{eqnarray}
  \dot\Omega_{\rm LT} &=& 68.5\ {\rm mas\ yr^{-1}} \\
  \nonumber\\
   \dot\omega_{\rm LT}&=& 0.
\end{eqnarray}
which correspond to  a shift $\Delta\nu$ of the  cross-track component  of the planetocentric position  \citep{Cri}   %
\begin{eqnarray}
  \Delta \nu_{\rm LT} &=& a\sqrt{1+\rp{e^2}{2}}\sin i\Delta\Omega_{\rm LT} = 572\ {\rm m}\ (T=1\ {\rm yr})
  %
  %
\end{eqnarray}
over the entire duration of the mission.
A total accuracy of the order of 1-10 m with respect to the km-level of the past Jupiter missions in reconstructing the Juno's orbit in a planetocentric frame does not seem
an unrealistic target, although much work is clearly required in order to have a more firm answer. Note that a 1-10 m accuracy implies a $0.2-2\%$ error in measuring the gravito-magnetic shift

The fact that the possibility of detecting the Lense-Thirring effect with Juno's orbit  seems worth of further consideration
can be preliminarily shown also with a different approach  with respect to the cumulative measurement over the full mission duration previously outlined.
Indeed, a gravity-science pass for Juno is defined by a continuous, coherent Doppler range-rate measurement plus and minus three hours of closest approach; in practice, most of the Lense-Thirring precession takes place just during such a six-hours pass, a near optimum condition. Another crucial factor is the orientation of the Earth to the Juno's orbit: our planet will be aligned 67 deg from the probe's orbital plane at approximately two degrees south latitude on the jovian equator. Preliminary numerical simulations of the Juno's Lense-Thirring Doppler range-rate signal show that such an orbital geometry represent a perfect compromise for measuring both the Jupiter's even zonal harmonics and the gravitomagnetic  signal itself. Indeed, it turns out that the maximum Lense-Thirring Doppler signal over a single six-hour gravity pass is of the order of hundred $\mu$m s$^{-1}$, while the limit of accuracy for Juno's Ka-band Doppler system is about one  $\mu$m s$^{-1}$ over such a pass. Thus, even by taking 25 repeated passes, over a total of approximately 33, it would be possible to reach a measurement precision below the percent level.

Concerning the magnitude of the jovian gravito-magnetic field, it must be noted that in literature there are estimates for $S$ which point towards smaller values than \rfr{LT} by a factor $1.5-1.6$; for example, \citet{Mac08}
yield \eqi S = 4.14 \times 10^{38}\ {\rm kg\ m^2\ s^{-1}};\lb{SJap} \eqf
the ratio of \rfr{LT} to \rfr{SJap}, i.e. 1.6, is close to 1.5 coming from the ratio  of $C/MR^2=2/5=0.4$, valid for a homogenous sphere, to $C/MR^2=0.264$ by \citet{Irw03} who assumes a concentration of mass towards the Jupiter's center.
Here we will  consider only the systematic uncertainty induced by the imperfect knowledge of the Newtonian part of the Jupiter's gravitational field; we will use \rfr{LT} for $S$.

\section{The systematic uncertainty due to the zonal harmonics of the Jupiter's gravitational potential}
A major source of systematic uncertainty is represented, in principle, by the departures of the Jupiter's gravitational field from spherical symmetry \citep{And76}.
\subsection{Analytical calculations}
Indeed,
the  zonal ($m=0$) harmonic coefficients $J_{\ell}$ of the multipolar expansion of the Newtonian part of the planet's gravitational potential give rise to long-period, that is averaged over one orbital revolution, orbital effects  on the longitude of the ascending node $\Omega$, the argument of pericentre $\omega$ and the mean anomaly $\mathcal{M}$  \citep{Kau}
\begin{eqnarray}
  \left\langle\dot\Omega\right\rangle &=& \sum_{\ell=2}\dot\Omega_{.\ell} J_{\ell}, \\
  \nonumber\\
  \left\langle\dot\omega\right\rangle &=& \sum_{\ell=2}\dot\omega_{.\ell} J_{\ell}, \\
  \nonumber\\
  \left\langle\dot{\mathcal{M}}\right\rangle &=& \sum_{\ell=2}\dot{\mathcal{M}}_{.\ell} J_{\ell},\\
\end{eqnarray}
where $\dot\Omega_{.\ell},\dot\omega_{.\ell},\dot{\mathcal{M}}_{.\ell}$ are coefficients depending on the planet's $GM$ and equatorial radius $R$, and on the spacecraft's inclination $i$ and eccentricity $e$ through the inclination $F_{\ell m p}(i)$ and eccentricity $G_{\ell p q}(e)$ functions, respectively \citep{Kau}.
Note that one of the major scientific goals of the Juno mission is  a greatly improved determination of just the harmonic coefficients of the jovian gravity potential; for the present-day values of the zonals\footnote{The Jupiter gravity field is essentially determined by the Pioneer 11 flyby at 1.6$R_{\rm Jup}$ \citep{And76}; Voyager added little, and Galileo, which never got close to Jupiter, added nothing.} for $\ell=2,3,4,6$ see Table \ref{zonals}.
\begin{table}[!h]
\caption{\footnotesize{Zonal harmonics of the Jupiter's gravity field according to the JUP230 orbit solution \protect\citep{JUP230}.}}\label{zonals}
\centering
\bigskip
\begin{tabular}{cccc}
\hline\noalign{\smallskip}
$J_2\ (\times 10^6)$ & $J_3\ (\times 10^6)$ & $J_4\ (\times 10^6)$  & $J_6\ (\times 10^6)$\\
\noalign{\smallskip}\hline\noalign{\smallskip}
$14696.43\pm 0.21$ & $-0.64\pm 0.90$ & $-587.14\pm 1.68$  & $34.25\pm 5.22$ \\
\noalign{\smallskip}\hline
\end{tabular}
\end{table}
According to \citet{And04}, it might be possible to determine the first three even zonals with an accuracy of $10^{-9}$ and the other ones up  to $\ell=30$  at a $10^{-8}$ level. Concerning $J_2$, this would be an improvement of two orders of magnitude with respect to Table \ref{zonals}, while the improvements in $J_4$ and $J_6$ would be of the order of three orders of magnitude.
By using the results we are going to present below for the long-period node and pericenter precessions, it can be shown that determining the low degree zonals at $10^{-9}$ level of accuracy translates into an accuracy of the order of $0.5-1$ mas yr$^{-1}$ in $\dot\Omega$ and $\dot\omega$, thus confirming the expectations of the previous Section.

For the long-period terms
 the condition
\eqi \ell-2p+q=0\eqf is fulfilled. Thus,
{\small\begin{eqnarray}
  \dot\Omega_{.\ell} &=& -\rp{n}{\sqrt{1-e^2}\sin i}\left(\rp{R}{a}\right)^{\ell}\sum_{p=0}^{\ell}\left[\dert{F_{\ell 0 p}}{i}G_{\ell p (2p-\ell)}W_{\ell 0 p(2p-\ell)}\right], \\
  \nonumber\\
  \dot\omega_{.\ell} &=& -\rp{n}{\sqrt{1-e^2}}\left(\rp{R}{a}\right)^{\ell}\sum_{p=0}^{\ell}\left[-\cot i\dert{F_{\ell 0 p}}{i}G_{\ell p (2p-\ell)}+
  \rp{(1-e^2)}{e}F_{\ell 0 p}\dert{G_{\ell p(2p-\ell)}}{e}\right]W_{\ell 0 p(2p-\ell)}, \\
  \nonumber\\
  \dot{\mathcal{M}}_{.\ell}  &=& n \left\{1 - \left(\rp{R}{a}\right)^{\ell}\sum_{p=0}^{\ell}F_{\ell 0 p}\left[6G_{\ell p(2p-\ell)} -\rp{(1-e^2)}{e}\dert{G_{\ell p(2p-\ell)}}{e}\right]W_{\ell 0 p(2p-\ell)}\right\},
\end{eqnarray}
}
where $W_{\ell 0p(2p-\ell)}$ are trigonometric functions having the pericentre as their argument
and $n=\sqrt{GM/a^3}$ is the unperturbed Keplerian mean motion.
Contrary to the case of small eccentricity satellites, in this case
we will be forced to keep all the terms of order $\mathcal{O}(e^k)$ with $k>2$ in computing the eccentricity functions for given pairs of $\ell$ and $p$. Moreover, we will need also
all the non-zero eccentricity and inclination functions  for a given degree $\ell$, that is we will consider all the non-vanishing terms with $0\leq p\leq \ell$. First, we will extend our calculations to the  even zonals so far determined, i.e.  $\ell=2,4,6$.
In this case,
\eqi W_{\ell 0 p(2p-\ell)}=\cos[(\ell -2p)\dot\omega t]=\cos(q\dot\omega t).\eqf
It must be noted that the terms with
\eqi p = \rp{\ell}{2},\ q=0\eqf yield secular precessions, i.e. $W_{\ell 0 \rp{\ell}{2} 0}=1$, while those with
\eqi q=2p-\ell\neq 0\eqf induces harmonic signals with circular frequencies $-q\dot\omega$.

For the degree $\ell=2$ the non-vanishing inclination and eccentricity functions and their derivatives are
\eqi F_{201} = \rp{3}{4}\sin^2 i - \rp{1}{2}.\eqf

\eqi \dert{F_{201}}{i} = \rp{3}{2}\sin i\cos i.\eqf

\eqi G_{210}=\rp{1}{(1-e^2)^{3/2}}.\eqf

\eqi\dert{G_{210}}{e} = \rp{2}{3}\rp{e}{(1-e^2)^{4/3}}.\eqf

In this case, only secular precessions occur.

For $\ell=4$ we have
\begin{eqnarray}
  F_{401} &=&  -\rp{35}{32}\sin^4 i + \rp{15}{16}\sin ^2 i   \ = \ F_{403},\\
  \nonumber\\
  F_{402} &=& \rp{105}{64}\sin^4 i - \rp{15}{8}\sin ^2 i + \rp{3}{8}.
\end{eqnarray}

\begin{eqnarray}
  \dert{F_{401}}{i} &=& \left(-\rp{35}{8}\sin^3 i + \rp{15}{8}\sin i\right)\cos i  \ = \ \dert{F_{403}}{i},\\
  \nonumber\\
  \dert{F_{402}}{i} &=& \left(\rp{105}{16}\sin^3 i - \rp{15}{4}\sin i\right)\cos i.
\end{eqnarray}

\begin{eqnarray}
  G_{41-2} &=& \rp{3}{4}\rp{e^2}{(1-e^2)^{7/2}} \ = \ G_{432},\\
  \nonumber\\
  G_{420} &=& \rp{1+\rp{3}{2}e^2}{(1-e^2)^{7/2}}.
\end{eqnarray}

\begin{eqnarray}
  \dert{G_{41-2}}{e} &=&  \rp{3}{14}\rp{e\left(7-6e^2\right)}{(1-e^2)^{8/7}}\ = \  \dert{G_{432}}{e},\\
  \nonumber\\
   \dert{G_{420}}{e} &=& \rp{1}{7}\rp{e\left(23-18e^2\right)}{(1-e^2)^{8/7}}
\end{eqnarray}
In addition to secular terms, also harmonic signals with the frequencies $\pm 2\dot\omega$ are present.

In the case of $\ell = 6$ the inclination and eccentricity functions, along with their derivatives, are
\begin{eqnarray}
  F_{601} &=& \rp{693}{512}\sin^6 i - \rp{315}{256}\sin^4 i \ = \ F_{605}, \\
  \nonumber\\
  F_{602} &=& -\rp{3465}{1024}\sin^6 i + \rp{315}{64}\sin^4 i - \rp{105}{64}\sin ^2 i\ = \ F_{604}, \\
  \nonumber\\
  F_{603} &=& \rp{1155}{256}\sin^6 i - \rp{945}{64}\sin^4 i + \rp{105}{32}\sin ^2 i -\rp{5}{16}.
\end{eqnarray}

\begin{eqnarray}
  \dert{F_{601}}{i} &=& \left(\rp{2079}{256}\sin^5 i - \rp{315}{64}\sin^3 i\right)\cos i\ =\   \dert{F_{605}}{i},\\
  \nonumber\\
  \dert{F_{602}}{i}  &=& \left(-\rp{10395}{512}\sin^5 i + \rp{315}{16}\sin^3 i - \rp{105}{32}\sin  i\right)\cos i \ =\   \dert{F_{604}}{i}, \\
  \nonumber\\
  \dert{F_{603}}{i}  &=& \left(\rp{3465}{128}\sin^5 i - \rp{945}{16}\sin^3 i + \rp{105}{16}\sin  i\right)\cos i.
\end{eqnarray}

\begin{eqnarray}
  G_{61-4} &=& \rp{5}{16}\rp{e^4}{(1-e^2)^{11/2}}   \ = \ G_{654},\\
  \nonumber\\
  G_{62-2} &=& \rp{5}{2}\rp{e^2}{(1-e^2)^{11/2}}\left(1 + \rp{e^2}{8}\right)  \ = \ G_{642},\\
  \nonumber\\
  G_{630} &=& \rp{1}{(1-e^2)^{11/2}}\left( 1 + 5e^2 + \rp{15}{8}e^4 \right) .
\end{eqnarray}

\begin{eqnarray}
\dert{G_{61-4}}{e} & = &\rp{5}{4}\rp{e^3\left(1-\rp{21}{22}e^2\right)}{(1-e^2)^{12/11}}\ = \ \dert{G_{654}}{e},\\
 \nonumber\\
\dert{G_{62-2}}{e} & = &  5\rp{e\left(1-\rp{29}{44}e^2 - \rp{21}{88}e^4\right)}{(1-e^2)^{12/11}} \ = \ \dert{G_{642}}{e},\\
\nonumber\\
\dert{G_{630}}{e} &=& \rp{7}{44}\rp{e\left(64-10e^2 - 45 e^4\right)}{(1-e^2)^{12/11}}.
\end{eqnarray}
In addition to the secular rates, also harmonic signals with frequencies $\pm 4\dot\omega,\pm 2\dot\omega$ are present.

Let us, now, focus on the action of the odd ($\ell=3,5,7,...$) zonal ($m=0$) harmonics.
In this case,
\eqi W_{\ell 0 p(2p-\ell)}=\sin[(\ell-2p)\dot\omega t]=-\sin(q\dot\omega t),\eqf
 so that only harmonic terms occur for $q\neq 0$.

For $\ell=3$ we have

\begin{eqnarray}
F_{301} &=&  \rp{15}{16}\sin^3 i -\rp{3}{4}\sin i\ =\ -F_{302}, \\
\nonumber\\
\dert{F_{301}}{i} & = & \left( \rp{45}{16}\sin^2 i -\rp{3}{4}\right)\cos i = -\dert{F_{302}}{i}.
\end{eqnarray}

\begin{eqnarray}
  G_{31-1} &=& \rp{e}{(1-e^2)^{5/2}}\ =\ G_{321}, \\
  \nonumber\\
  \dert{G_{31-1}}{e} &=&  \rp{1-\rp{3}{5}e^2}{(1-e^2)^{6/5}}\ =\ \dert{G_{321}}{e}.
\end{eqnarray}

Thus, we have long-period effects varying with the frequencies $\pm\dot\omega$.

Concerning the even and odd zonal long-period harmonic terms, it must be noted that  they can be approximated by secular precessions with a high level of accuracy over the expected 1-yr lifetime $T$ of  Juno because the period of its periJove is of the order of $\approx 500$ yr, i.e.
\begin{eqnarray}
\cos (q\omega) & = & \cos(q\dot\omega t + q\omega_0)\approx\cos(q\omega_0),\ 0\leq t\leq T,\\
 \nonumber\\
 \sin\omega   & = & \sin(\dot\omega t + \omega_0)\approx\sin\omega_0,\ 0\leq t\leq T.
 \end{eqnarray}
 Thus, the choice of the initial condition $\omega_0$ will be crucial in determining their impact.

Now,  it would be possible, in principle, to use the node of Juno to measure the gravito-magnetic effect. Indeed, the Lense-Thirring node precession is independent of $i$, while all the zonal precessions  of $\Omega$ vanish for $i=90$ deg. It does not occur for the periJove and the mean anomaly, but they are not affected by the gravito-magnetic force for $i=90$ deg. In reality, the situation will be likely different because of the unavoidable orbit injection errors which will induce some departures $\delta i$ of the Juno's orbital plane from the ideal polar configuration. Thus, unwanted, corrupting node zonal secular precessions will appear; their mis-modeling due to the uncertainties $\delta J_{\ell}$  may swamp the recovery of the Lense-Thirring effect if their determination by Juno will not be accurate enough.
Note that there is no risk of some sort of a-priori `imprint' effect of the Lense-Thirring effect itself on the values of the zonals retrieved from the Juno's periJove motion because the gravito-magnetic pericenter precession vanishes for polar orbits.

\begin{figure}
\includegraphics[width=\columnwidth]{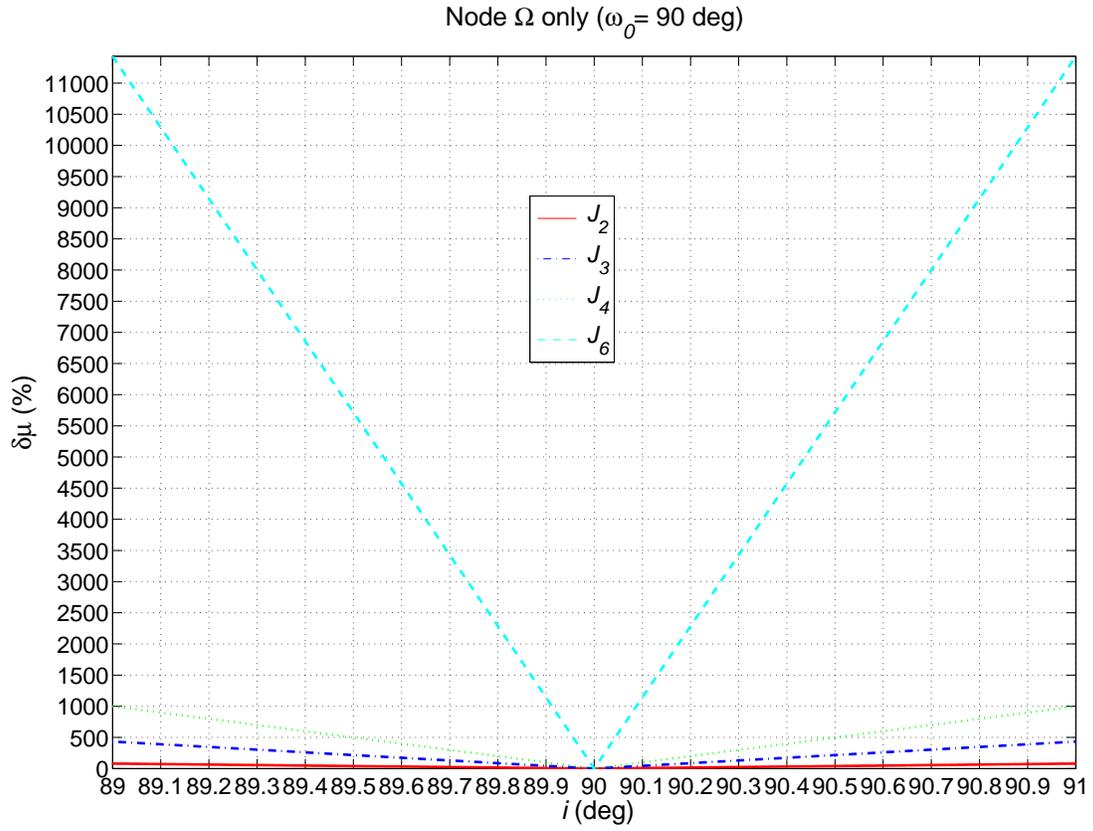}
 \caption{\footnotesize{\label{JUNO_node} Systematic percent bias on the Lense-Thirring node precession induced by the mis-modeling in the zonals $J_2,J_3,J_4,J_6$ according to Table \ref{zonals} for $89$ deg $\leq i\leq 91$ deg and $\omega_0=90$ deg.} }
\end{figure}

By assuming the values quoted in Table \ref{zonals} for the uncertainties $\delta J_{\ell}$, $\ell=2,3,4,6$, let us see what is the impact of a non-perfectly  polar orbital geometry on the node Lense-Thirring precessions.  The results are depicted in Figure \ref{JUNO_node} for each degree $\ell$ separately; the initial condition $\omega_0=90$ deg has been used.  It should be noted that, in view of the likely correlations among the determined zonals, a realistic upper bound of the total bias due to them can be computed by taking the linear sum of each mis-modeled terms.
The major source of bias is the so far poorly known $J_6$; an improvement of four orders of magnitude, which sounds rather unlikely to be obtained even with Juno \citep{And04}, would be required to push its aliasing effect at a percent level of the Lense-Thirring effect. The situation for the other zonals is more favorable; $J_4$ should be known better than now by a factor 1000, which is, instead, a realistic goal according to \citet{And04}.
Thus, we conclude that a nearly-polar orbit 1 deg off the ideal 90 deg case would likely prevent to obtain a measurement of the gravito-magnetic node precession at a decent level of accuracy.

Thanks to the high eccentricity of the Juno's orbit, also the periJove and the mean anomaly are well defined, so that they can be used in a suitable way to remove the bias of $J_6$ and $J_2$.
Let us write down
\begin{eqnarray}
  \delta\dot\Omega &=& \dot\Omega_{.2}\delta J_2 + \dot\Omega_{.6}\delta J_6 + \mu\dot\Omega_{\rm LT} + \Delta_{\Omega}, \\
  \nonumber\\
    \delta\dot\omega &=& \dot\omega_{.2}\delta J_2 + \dot\omega_{.6}\delta J_6 + \mu\dot\omega_{\rm LT} + \Delta_{\omega},  \\
    \nonumber\\
\delta\dot{\mathcal{M}} &=& \dot{\mathcal{M}}_{.2}\delta J_2 + \dot{\mathcal{M}}_{.6}\delta J_6 + \Delta_{\mathcal{M}}.
\end{eqnarray}
Here $\delta\dot\Psi$ denotes some sort of Observed-minus-Calculated ($O-C$) quantity for the rate of the Keplerian element $\Psi$ which accounts for every unmodeled/mis-modeled dynamical effects; it may be, for example, a correction to the modeled precessions to be phenomenologically estimated as a solve-for parameter of a global fit of the Juno's data, or it could be a computed time-series of\footnote{Since the Keplerian elements are not directly measurable quantities, we use here the term ``residual'' in an improper sense.}  ``residuals'' of $\Psi$ by suitably overlapping orbital arcs.  The gravito-magnetic force should be purposely not modeled in order to be fully present in $\delta\dot\Psi$.
The parameter $\mu$ is\footnote{It is not one of the standard PPN parameters, but it can be expressed in terms of $\gamma$ as $\mu=(1+\gamma)/2$.} 1 in GTR and 0 in Newtonian mechanics and accounts for the Lense-Thirring effect. The $\Delta$ terms include all the other systematic errors like the precessions induced by the mis-modeled parts of the second even zonal harmonic $\delta J_4$ and the first odd zonal harmonic $\delta J_3$,
the mis-modeling due to the uncertainty in Jupiter's $GM$, etc.
By solving for $\mu$ one obtains
\eqi \delta\dot\Omega + c_1\delta\dot\omega + c_2\delta\dot{\mathcal{M}}=\dot\Omega_{\rm LT} + c_1\dot\omega_{\rm LT}+\Delta,\lb{comb}\eqf
with
\begin{eqnarray}
  c_1 &=& \rp{ {\dot{\mathcal{M}}_{.6}}\ {\dot\Omega_{.2}} - {\dot\Omega_{.6}}\ {\dot{\mathcal{M}}_{.2}}} {{\dot\omega_{.6}}\ {\dot{\mathcal{M}}_{.2}} -{\dot{\mathcal{M}}_{.6}}\ {\dot\omega_{.2}} }, \lb{coef1}\\
  \nonumber\\
  c_2 &=& \rp{ {\dot\Omega_{.6}}\ {\dot\omega_{.2}} - {\dot\omega_{.6}}\ {\dot\Omega_{.2}}} {{\dot\omega_{.6}}\ {\dot{\mathcal{M}}_{.2}} -{\dot{\mathcal{M}}_{.6}}\ {\dot\omega_{.2}} }. \lb{coef2}\end{eqnarray}
The combination of \rfr{comb}, with \rfr{coef1} and \rfr{coef2}, is designed, by construction, to single out the combined Lense-Thirring precessions and to cancel out the combined secular\footnote{We include in them also the long-period harmonic terms for the reasons explained before.} precessions due to $J_2$ and $J_6$ along with their mis-modeling. Instead, it is affected by $\Delta$ which acts as a systematic bias on the Lense-Thirring signal of interest. $\Delta$ globally includes the mis-modeled part of the combined precessions induced by $J_3$ and $J_4$; the sources of uncertainty reside in $J_3$ and $J_4$ themselves and in the Jupiter's $GM$ through the mean motion $n$ which enters the coefficients $\dot\Omega_{.\ell},\dot\omega_{.\ell},\dot{\mathcal{M}}_{.\ell}$.

\begin{figure}
\includegraphics[width=\columnwidth]{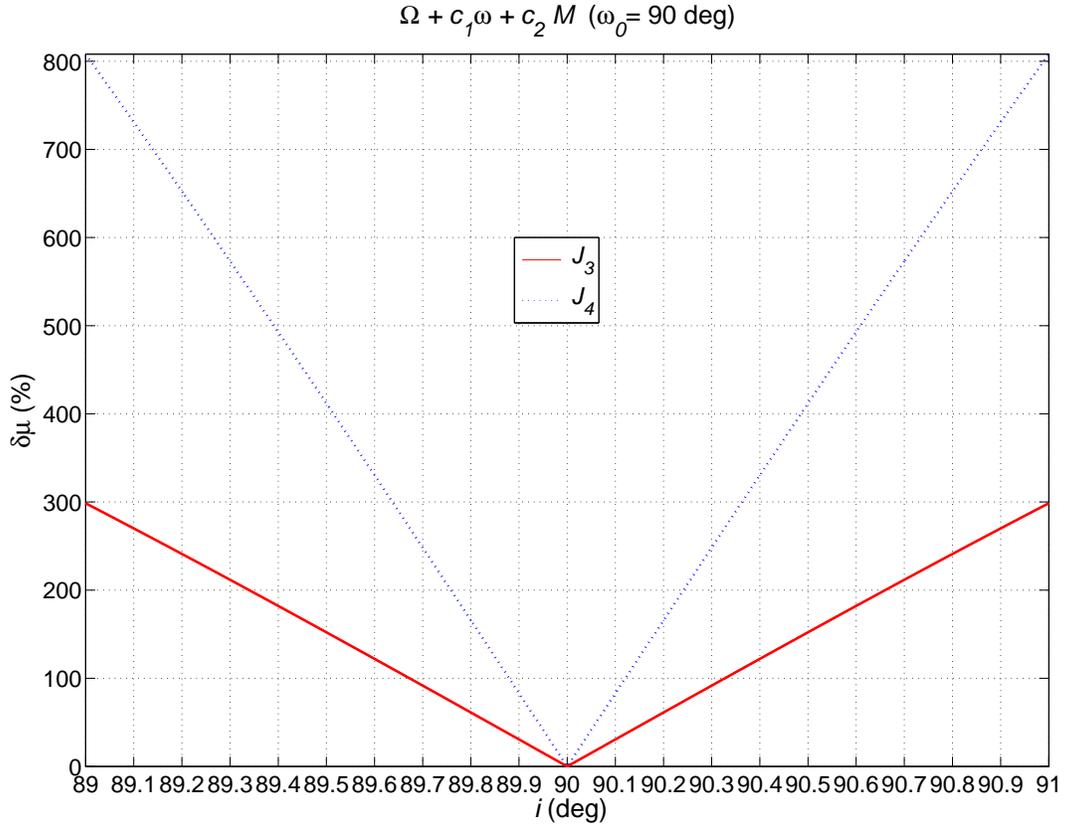}
 \caption{\footnotesize{\label{JUNO_combi_90} Systematic percent bias on the Lense-Thirring precessions, combined according to \rfr{comb}, induced by the mis-modeling in the uncanceled zonals $J_3,J_4$  (Table \ref{zonals}) for $89$ deg $\leq i\leq 91$ deg and $\omega_0=90$ deg. }}
\end{figure}

\begin{figure}
\includegraphics[width=\columnwidth]{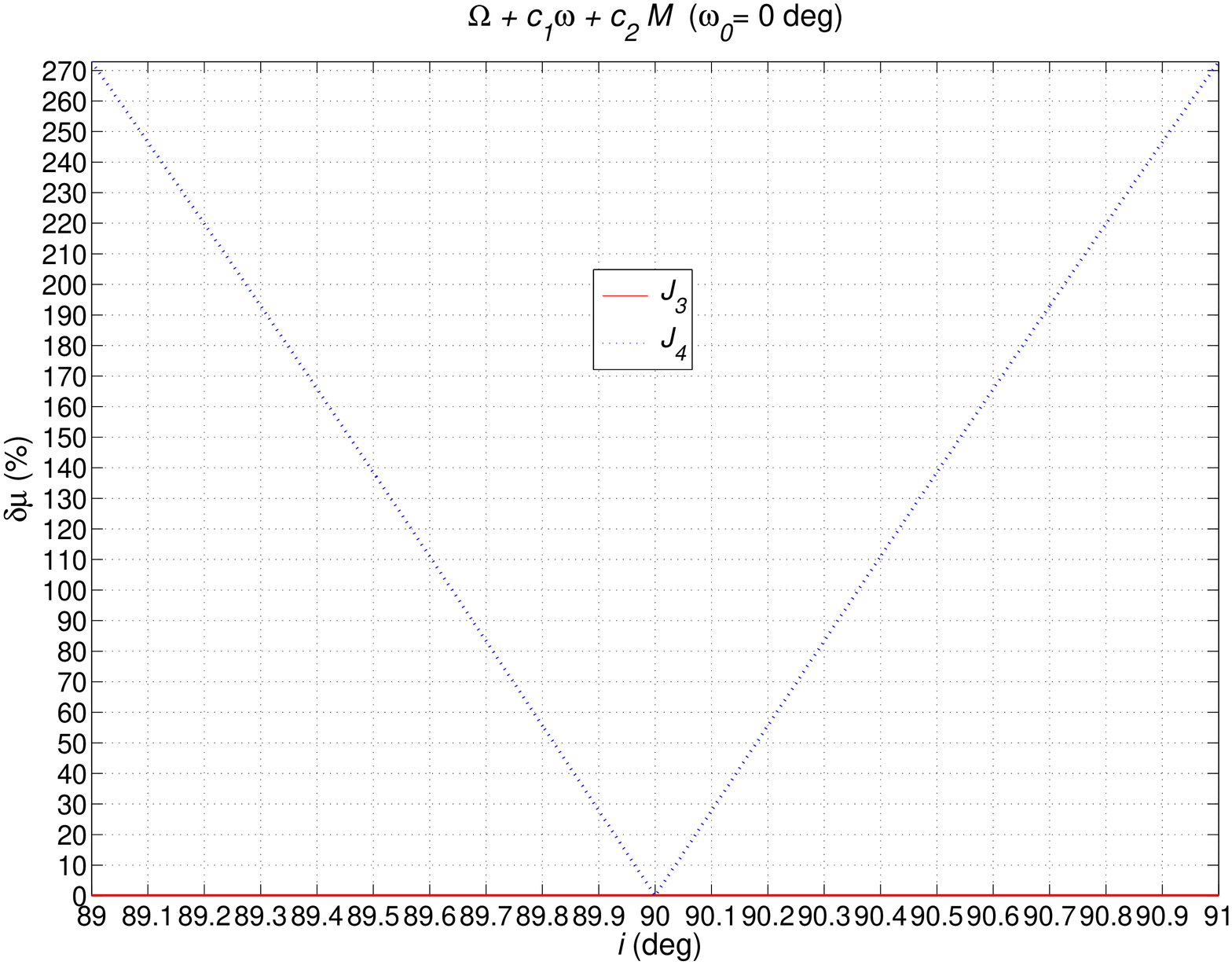}
 \caption{\footnotesize{\label{JUNO_combi_0} Systematic percent bias on the Lense-Thirring precessions, combined according to \rfr{comb}, induced by the mis-modeling in the uncanceled zonals $J_3,J_4$  (Table \ref{zonals}) for $89$ deg $\leq i\leq 91$ deg and $\omega_0=0$ deg. }}
\end{figure}

In Figure \ref{JUNO_combi_90} the impact of the mis-modeling in $J_3$ and $J_4$ for $\omega_0=90$ deg is depicted.
In Figure  \ref{JUNO_combi_0} we use $\omega_0 = 0$ deg.
In this case the situation is much more favorable because for a total departure of $\pm 1$ deg from $i=90$ deg, an improvement of only two orders of magnitude in $J_3$, which is, today, still compatible with zero, and $J_4$ would be needed to reach the percent level;
let us recall that the expected improvement in $J_4$ with respect to the results by \citet{JUP230} is of three orders of magnitude \citep{And04}.
Note that a value of $\omega_0$ far from 90 deg is preferable to minimize
the perturbations.

Another potential source of systematic error is the Jupiter's $GM$ whose uncertainty $\delta(GM)$ indirectly affects \rfr{comb} through the Keplerian mean motion $n$ entering the uncanceled $J_3$ and $J_4$ combined precessions; $\delta n$ is also present via the mean anomaly itself. However, it turns out that it is of no concern because, according to the present-day level of relative uncertainty  \citep{JUP230}
\eqi\rp{\delta(GM)}{GM}=1.6\times 10^{-8},\eqf
its impact on the combined Lense-Thirring precessions is well below the percent level.
\subsection{A numerical approach}
Also in this case, we  followed an alternative approach based on preliminary numerical simulations. We investigated the impact of the uncertainties in the first two jovian even zonals on a Juno's single six-hours pass by numerically simulating the probe's Doppler range-rate signals due to $\delta J_2$ and $\delta J_4$. By assuming for them values as large as $2\times 10^{-10}$ and $3\times 10^{-10}$, respectively, it turns out that the maximum Doppler shifts are roughly $1-1.5$ $\mu$m s$^{-1}$. Moreover, and more importantly, the time-dependent patterns of the even zonals' Doppler signals are quite different from the Lense-Thirring one removing the risk of an insidious mimicking bias. Another encouraging fact is that such simulations  indicate that an inclination of even 91 deg would not compromise the recovery of the gravitomagnetic signal of interest.
\section{Discussion and conclusions}
The node $\Omega$ of Juno, a recently approved spacecraft aimed to orbit Jupiter along a highly eccentric ($r_{\rm min}=1.06R_{\rm Jup}$, $r_{\rm max}=39R_{\rm Jup}$), polar ($i=90$ deg) trajectory during one year to accurately map, among other things, its gravitational field, will be displaced by the general relativistic gravito-magnetic Lense-Thirring effect by about 572 m over the entire duration of the mission.

We, first, explored the possibility of a high accuracy measurement of such an effect by performing  analytical calculations and interpreting them in a rather conservative fashion.
A meter-level accuracy in determining the jovicentric orbit of Juno should not be an unrealistic goal to be reached. Equivalently, the gravito-magnetic node precession of Juno amounts to 68.5 mas yr$^{-1}$, while the accuracy in measuring its node and periJove precessions should be of the order of $0.5-1$ mas yr$^{-1}$, given the expected improvements in our knowledge of the departure of the jovian gravitational field from spherical symmetry. If the Juno's orbit was perfectly polar, the long-period node precessions induced by the zonal harmonics $J_{\ell}, \ell=2,3,4,6,...$ of the non-spherical jovian gravitational potential would vanish, thus removing a major source of systematic alias on the Lense-Thirring secular precession. In reality, unavoidable orbit injection errors will displace the orbital plane of Juno from the ideal polar geometry; as a consequence, the mis-modeled part of the node zonal precessions would overwhelm the relativistic signal  for just $\delta i = \pm 1$ deg, in spite of the expected  improvements in $J_{\ell}, \ell=2,4,6$ of three orders of magnitude. A suitable linear combination of the node, the periJove $\omega$ and the mean anomaly $\mathcal{M}$ will allow to cancel out the effect of $J_2$ and $J_6$; the remaining uncanceled $J_3$ and $J_4$ will have an impact on the Lense-Thirring combined precessions which should be reduced down to the percent level or better by the improved low-degree zonals.

Instead of looking at the cumulative, secular effects over the entire duration of the mission, we also followed an alternative approach by looking at single Doppler range-rate  measurements over time spans six hours long centered on the the probe's closest approaches to Jupiter; it turned out that, in this way, the perspectives are even more favorable. We numerically simulated the characteristic Lense-Thirring pattern for a single science pass by finding a maximum value of the order of hundreds $\mu$m s$^{-1}$, while the expected precision level in Juno's Doppler measurements is of the order of one  $\mu$m s$^{-1}$. Thus, by exploiting about 25 of the planned 33 total passes of the mission it would be possible to reach a measurement accuracy below the percent level. We repeated our numerical analysis also for the Doppler range-rate signals of $J_2$ and $J_4$ by finding quite different patterns with respect to the gravito-magnetic one; moreover, for a level of mismodeling of the order of $2-3\times 10^{-10}$ in such zonals the maximum value of their biasing Doppler signals is about $1-1.5$ $\mu$m s$^{-1}$. Our numerical analysis also shows that a departure from the nominal polar orbital geometry as large as 1 deg would not compromise the successful outcome of the measurement of interest, contrary to the conservative conclusions of our analytical analysis. Thus, this approach shows that there is not a high correlation between the Lense-Thirring parameter and the jovian gravity field parameters, although a covariance analysis would be needed to prove it. However, such a covariance analysis is outside the scope of the present paper.

In conclusion, the potential error in the proposed Juno Lense-Thirring measurement is between $0.2$ and $5$ percent. Conversely, if one assumes the existence of gravito-magnetism as predicted by general relativity, the proposed measurement can also be considered as a direct, dynamical determination of the jovian proper angular momentum $S$ by means of  the Lense-Thirring effect at the percent level.

\section*{Acknowledgments}
I gratefully thank an anonymous referee for her/his remarkable efforts to improve the manuscript with important insights about the numerical simulations. I am also grateful to  J.D. Anderson for useful discussions.


\end{document}